\documentclass[a4paper,twoside]{article}

\usepackage{epsfig}
\usepackage{subcaption}
\usepackage{calc}
\usepackage{amssymb}
\usepackage{amstext}
\usepackage{amsmath}
\usepackage{amsthm}
\usepackage{multicol}
\usepackage{pslatex}
\usepackage{apalike}
\usepackage{algorithm2e}
\usepackage[bottom]{footmisc}

\usepackage{multirow}
\usepackage{tabularx}
\usepackage{bm}
\usepackage{graphicx}
\usepackage{upgreek}
\newcommand\inlineeqno{\stepcounter{equation}\ (\theequation)}
\usepackage{cancel}
\usepackage{balance}

\usepackage{SCITEPRESS}     

\begin{document}

\title{Recursive Gaussian Process Regression with Integrated Monotonicity Assumptions for Control Applications}


\author{\authorname{Ricus Husmann, Sven Weishaupt and Harald Aschemann}
\affiliation{Chair of Mechatronics, University of Rostock, Rostock, Germany}
\email{\{ricus.husmann, sven.weishaupt, harald.aschemann\}@uni-rostock.de}
}


\abstract{In this paper, we present an extension to the recursive Gaussian Process (RGP) regression that enables the satisfaction of inequality constraints and is well suited for a real-time execution in control applications. The soft inequality constraints are integrated by introducing an additional extended Kalman Filter (EKF) update step using pseudo-measurements. The sequential formulation of the algorithm and several developed heuristics ensure both the performance and a low computational effort of the algorithm. A special focus lies on an efficient consideration of monotonicity assumptions for GPs in the form of inequality constraints. The algorithm is statistically validated in simulations, where the possible advantages in comparison with the standard RGP algorithm become obvious. The paper is concluded with a successful experimental validation of the developed algorithm for the monotonicity-preserving learning of heat transfer values for the control of a vapor compression cycle evaporator, leveraging a previously published partial input output linearization (IOL).}

\onecolumn \maketitle \normalsize \setcounter{footnote}{0} \vfill

\section{Introduction}
A lot of progress can be noticed regarding the online identification of system models. An example for a basic method is given by  the definition of parametric functions like polynomial ansatz functions and the subsequent use of a recursive least-squares regression as described in \cite{Blum:1957}. If models with non-measurable states or parameters are to be identified, this method can be extended to linear Kalman Filters (KF) or Unscented/Extended Kalman Filters (UKF/EKF) as proposed in \cite{Kalman:1960}, \cite{Julier:1997}. In the presence of additional inequality constraints, especially Moving Horizon Estimation (MHE) techniques are suitable \cite{Haseltine:2005}. If a more general approach is envisaged, there exist online-capable training methods of neural networks as discussed in \cite{Jain:2014}.

Gaussian Processes (GP) have been established as a popular non-parametric alternative to neural networks (NNs). They are usually more data-efficient than neural networks, robust to overfitting and -- as main advantage in comparison to NNs -- provide an uncertainty quantification for the predicted value, see \cite{Schuerch:2020}. The non-parametric character and the $O(n^3)$ rise of computational load in dependency of the utilized data points, however, poses a big challenge for their online implementation. Nevertheless, the literature offers suitable methods to address this problem, like active-set methods as described in \cite{Quinonero:2005}, which limit the number of utilized measurement points. Furthermore, a promising algorithm was presented by \cite{Huber:2013} with the recursive Gaussian Process regression (RGP). The idea here is to define the GPs as parametric functions based on user-defined basis vectors. This preserves many benefits of the GP regression while maintaining a small computational load.

For many modeling tasks, a certain previous knowledge is available. This may include bounds on model outputs or monotonicity assumptions. This knowledge might come from physical properties or in the form of stability-conserving constraints in a control setting. Nevertheless, the application of this knowledge during learning has the potential to provide superior models with far less data. In neural networks, such assumptions might be considered by an altering of the reward function as in \cite{Desai:2021}. For standard GPs, a number of methods are available for considering assumptions, e.g. regarding the system structure as presented in \cite{Beckers:2022} or for inequality constraints as shown in \cite{Veiga:2020}. To the best of our knowledge, however, the integration of inequality constraints for recursive GPs represents a new development. 

\newpage

In \cite{VCC_Part:2024}, a partial IOL for Vapor Compression Cycle (VCC) Control including an integral feedback regarding the heat transfer value was presented. As mentioned in that paper, the usage of the steady state of the integrator to derive a data-based model seems a promising idea. Since monotonicity assumptions hold for the input dependencies of such a heat transfer value model, this  application represents an interesting candidate for the validation of the RGP algorithm with the inclusion of monotonicity assumptions.

\vspace{0.2 cm}
\noindent The main contributions of the paper are:
\begin{itemize}
\item Computationally efficient consideration of inequality constraints w.r.t. Gaussian variables in a sequential EKF update
\item Integrating monotonicity assumptions into RGP regression by means of inequality constraints
\item Real-time implementation of the presented online-learning algorithm regarding a data driven RGP model for the heat-transfer value of a VCC
\end{itemize}
\vspace{0.2 cm}

The paper is structured as follows: First, we recapitulate the RGP in Sec.~\ref{sec:RGP}. Then, the generalized approach for considering inequality constraints is presented in Subsec.~\ref{sec:Ineq_Constr}. Afterwards, we introduce the special case of formulating GP monotonicity as such a constraint in Subsec.~\ref{sec:Monotonicity}, summarize and discuss the complete algorithm. As a first validation in Sec.~\ref{sec:simval}, we statistically analyze the validity of the algorithm for a simulation example. Finally, an experimental validation is provided in Sec.~\ref{sec:expval} with an implementation of the real-time algorithm on a VCC test rig. The paper finishes with a conclusion and an outlook.

\section{Recursive Gaussian Process Regression}
\label{sec:RGP}
In this chapter, we briefly describe recursive Gaussian Process regression (RGP) as presented in \cite{Huber:2013} and extended in \cite{Huber:2014}. As usual, we utilize a Squared Exponential (SE) kernel
\begin{equation}
k(\bm{x},\bm{x}')=\sigma_K^2 \cdot \exp(-(\bm{x}-\bm{x}')^T(\bm{x}-\bm{x}')(2 L)^{-1}) \,
\end{equation}
with a zero mean function $m(\bm{X})=0$. In our application, the hyperparameters $L$ and $\sigma_K$ are user-defined. As elaborated in Subsec.~\ref{sec:Inp_Norm}, a joint length scale $L$ is defined for all input dimensions, and the possibly different input ranges are addressed by an extra normalization step. For each GP, we consider only one measurement $y_k$ per time step $k$, with a corresponding measurement variance $\sigma_{y}^2$. As detailed in Subsec.~\ref{sec:Inp_Norm}, $\bm{X}$ refers to the constant basis vectors, which are defined during initialization, and $\bm{X}_k$ denotes the current test input. The mean values $\bm{\mu}^g_{k}$ and the covariance matrix $\bm{C}_{k}^g$ of the kernels, which are updated recursively, give the RGP algorithm a KF-like structure.

The following variables can be precalculated offline:
 \begin{center}
\begin{tabularx}{1.028\columnwidth}{l X r} 
  \multirow{4}{1em}{\rotatebox{90}{{Offline}}} & $\bm{K}=k(\bm{X},\bm{X})$,  &$\inlineeqno $ \\ 
& $\bm{K}_I=\bm{K}^{-1} $,& \\ 
& $\bm{\mu}_{0}^g= \bm{0}$ ,&\\ 
& $\bm{C}_{0}^g= \bm{K} \,.$ &\\  
\end{tabularx}
\end{center}
Given a zero mean function and a single measurement per time step, the inference step simplifies to:
\begin{center}
\begin{tabularx}{1.028\columnwidth}{l X r} 
  \multirow{3}{1em}{\rotatebox{90}{{Inference}}} & $\bm{J}_{k} =k(\bm{X}_k,\bm{X}) \cdot \bm{K}_I $  ,&$\inlineeqno$\\ 
& $\mu_{k}^p=\bm{J}_k \cdot \bm{\mu}^g_{k} $,& \\ 
& $C_{k}^p=\cancelto{\sigma_K^2}{k(\bm{X}_k,\bm{X}_k)}+\bm{J}_k (\bm{C}_{k}^g-\bm{K}) \bm{J}_k^T \,,$ & \\  
\end{tabularx}
\end{center}
where the superscript $p$ indicates the GP prediction for the test inputs $\bm{X}_k$. This prediction is used in the following update step:
\begin{center}
\begin{tabularx}{1.028\columnwidth}{l X r} 
  \multirow{3}{1em}{\rotatebox{90}{{Update}}} & $\bm{G}_{k}=\bm{C}_{k}^g \bm{J}_k^T \cdot(C_{k}^p+ \sigma_{y}^2 )^{-1}$ , &$\inlineeqno$ \\ 
& $\bm{\mu}_{k+1}^g= \bm{\mu}_{k}^g+\bm{G}_{k} \cdot (y_{k}- \mu_{k}^p) $,& \\ 
& $\bm{C}_{k+1}^g=\bm{C}_{k}^g-\bm{G}_{k} \bm{J}_k\bm{C}_{k}^g \,.$ & \\ 
\end{tabularx}
\end{center}

\subsection{Input Normalization} 
\label{sec:Inp_Norm}
We define the basis vectors for all input axes as an equidistant grid with step size $1$. This leads to a $\left(\prod_{i=1}^{n_{X}} N_i \right) \, \times \, n_{X}$-dimensional matrix which contains all vertices of the grid, where $n_{X}$ denotes the input dimension of the GP, and $N_i$ the number of points in the respective dimension. An example for the basis vectors for $n_X=2$ , $N_1=2$ and $N_2=3$ is
\begin{equation}
\bm{X}=\left[\begin{matrix}  \bm{X}_1^T\\ \bm{X}_2^T\end{matrix}\right]^T=\left[\begin{matrix}  0 &1 &0&1&0&1 \\  0 &0 &1&1&2&2 \end{matrix}\right]^T \,.
\end{equation}
To consider the ranges of the actual inputs $\zeta_i$ in each dimension, we introduce the normalization step
\begin{equation}
X_{i,k}=f_{norm}(\zeta_{k,i})=(\zeta_{i,k}-\underline{\zeta}_{i}) \cdot \underbrace{\frac{N_i-1}{\overline{\zeta}_{i}-\underline{\zeta}_{i}}}_{\beta_i}  \,,
\end{equation}
which is applied before each GP evaluation. Here  $\underline{\zeta}_{i}$ and $\overline{\zeta}_{i}$ denote the corresponding lower and upper bounds of the input, and $\beta_i$ is a constant factor, which is used in Subsec.~\ref{sec:Monotonicity}. 

The normalization and the use of a joint length $L$ was originally introduced to handle numerical issues that may occur for large $L$ in the standard inversion-based RGP formulation applied in this paper. With the normalization step, a universal maximum $L_{max}$ -- independent of the system -- can be determined to maintain numerical stability. Please note that an alternative decomposition-based online solution is available, which is described in \cite{RGP_dKF:2025}. An additional benefit, independent of the RGP formulation is given, however, by the reduction of free hyperparameters. Consequently, it has also been used in \cite{RGP_dKF:2025}.

\section{Monotonicity Constraint Integration}
In this chapter, we present our implementation to enforce (soft) monotonicity constraints for RGPs. The proposed method is based upon an EKF update for inequality constraints, which is described in the sequel after the precise problem formulation. Since the real-time capability is required, we put emphasis on a computational speedup of the algorithm by usage of reformulations and heuristics in the next subsections. Afterwards, we present the formulation of GP monotonicity as a constraint, summarize and discuss the complete algorithm.

We consider a hidden function ${y}_k=z_k(\bm{\zeta}_k)$ that is dependent on deterministic inputs $\bm{\zeta}_k$. The output ${y}_k$ of the hidden function $z_k$ is measurable, with zero-mean Gaussian measurement noise with a variance of $\sigma_y^2$. We assume previous knowledge of the monotonicity of  $z_k$ w.r.t. its inputs, which can be stated in an inequality constraint regarding the partial derivatives $\frac{\partial z_k}{\partial \zeta_i}\gtrless 0$. To enable safety margins, this is generalized to $\frac{\partial z_k}{\partial \zeta_i}\gtrless B_i$, where $B_i$ is a constant characterizing the boundary of the constraint.

\subsection{EKF Update for Inequality Constraints}
\label{sec:Ineq_Constr}
The direct consideration of hard inequality constraints on Gaussian variables leads to truncated Gaussians, see \cite{Tully:2011}. For univariate Gaussians, the resulting mean and covariance can be calculated efficiently. For multivariate Gaussians and inequality constraints that dependent on several Gaussian input variables, however, exact solutions usually necessitate numerical methods. Here, \cite{Simon:2006} provides an overview. \cite{Simon:2006} also discusses the use of equality constraints as exact pseudo-measurements within a KF update. This is related, however, to some numerical issues since exact measurements lead to rank-deficient updates in a KF. The alternative soft constraints, where the pseudo-measurement is considered with a small uncertainty, are not subject to this problem. In this paper, hence, we build upon this approach and extend it towards inequality constraints. 

The basic idea is to implement inequality constraints as pseudo-measurements using a ReLU measurement function as well as an EKF-update. In the case of an inactive inequality constraint in the current step, the ReLU function in combination with the EKF "hides" the inequality constraint in the update, otherwise it is considered as an equality constraint. Here, some parallels to the active-set method for constrained optimization become obvious, see \cite{Nocedal:2006}.  These parallels are for example also drawn in \cite{Gupta:2007}. There however in combination with, projection and gain-limiting methods instead of pseudo-measurements. Of course, a truncated Gaussian may differ quite dramatically in shape from a Gaussian distribution. As a results, this linearization-based approach may cause large errors in the covariance. To rule out over-approximation errors by this effect, the overall covariance update by the EKF inequality constraint is discarded at the end, as described later. This measure contributes to the "softness" of the constraints.

To simplify the implementation, we standardize all inequalities $j$ by means of the sign indicator variable $s_j$: $\hat{y}_{IC,1}<B_1 \equiv s_1\cdot (\hat{y}_{IC,1}-B_1)<0$ with $s_1=1$, $\hat{y}_{IC,2}>B_2\equiv s_2 \cdot (\hat{y}_{IC,2}-B_2)<0$ with $s_2=-1$. This corresponds to linear inequalities of the type $s_j (h_{IC,j}^T \bm{x}_k-B_j)<0$, where $\bm{x}_k$ denotes the state vector. Now, we introduce the nonlinear measurement function, which is evaluated with the mean values
\begin{equation}
\hat{y}_{IC,j}=\tilde{h}_{IC,j}(\bm{x}_k=\bm{\mu}_{k+1}^g)=\mathrm{ReLU} (s_j (h_{IC,j}^T\bm{\mu}_{k+1}^g-B_j))  \,.
\end{equation}
 Due to the standardization of the inequalities, all the pseudo-measurements become $y_{IC,j}=0$. The measurement functions $\bm{\tilde{h}}_{IC}=\left[{\tilde{h}}_{IC,1},{\tilde{h}}_{IC,2},..\right]^T$ can be concatenated in a vector. 

The partial derivative of the measurement, necessary for the EKF update, is given by
\begin{align}
\bm{\hat{h}}_{IC,j,k}^T = \left(\left.\frac{\partial \tilde{h}_{IC,j}(\bm{x}_k)}{\partial \bm{x}_k}\right|_{\bm{x}_k=\bm{\mu}_{k+1}^g}  \right)^T \nonumber \\ 
=\left\{ 
  \begin{array}{ c l }
    s_j  h_{IC,j}^T & \quad \mathrm{if}\quad s_j (h_{IC,j}^T\bm{\mu}_{k+1}^g-B_j)>0 \\
    \left[0,0,..\right]& \quad \mathrm{else}, \\
  \end{array}
\right.
\end{align}
which can be concatenated as well to the linearized measurement matrix $\bm{\hat{H}}_{IC,k}=\left[\bm{\hat{h}}_{IC,1,k},\bm{\hat{h}}_{IC,2,k},..\right]^T$. The update can then be computed according to a standard EKF, with the pseudo-measurements $y_{IC,j}=0$
\begin{align}
{\bm{\tilde{G}}}_{k}&=\bm{C}_{k+1}^g \bm{\hat{H}}_{IC,k}^T ({\bm{\hat{H}}_{IC,k}}  \bm{C}_{k+1}^g {\bm{\hat{H}}_{IC,k}}^T+ \bm{R}_{IC})^{-1} ~, \\  \nonumber
 \bm{\mu}_{k+1}^c&= \bm{\mu}_{k+1}^g-{\bm{\tilde{G}}}_{k} \cdot \bm{\tilde{h}}_{IC}(\bm{\mu}_{k+1}^g) ~, \\ \nonumber
 \bm{C}_{k+1}^c&=\bm{C}_{k+1}^g-{\bm{\tilde{G}}}_{k} \bm{\hat{H}}_{IC,k} \bm{C}_{k+1}^g, ~ 
\end{align}
where $\bm{R}_{IC}$ is the small diagonal pseudo-measurement noise matrix, with all $\bm{R}_{IC}(j,j)=r_{IC}$. The superscript $c$ denotes the constrained mean values and covariance.

\subsection{Reduction of the Computational Load}
\label{sec:seq_EKF}
Since we strive for real-time applicability of the algorithm, computational efficiency is crucial. Thus, instead of one batchwise EKF update, we perform $n_{IC}$ sequential EKF updates for each pseudo-measurement. The sequential KF is theoretically identical to the batchwise one -- as long as the (pseudo-)measurements are uncorrelated, see \cite{Simon2:2006}. This assumption is met in our case because $\bm{R}_{IC}$ is diagonal. For the EKF, additional conditions have to be taken into account as discussed in Subsec.~\ref{sec:discussion}. Please note that numerical differences between the batchwise and sequential KF update may occur for large dimensions $n_{IC}$. Consequently, the exact reformulation of the batchwise EKF update to a sequential EKF update can be stated as follows:\\
\\
Set $\bm{C}_{k+1,1}^c=\bm{C}_{k+1}^p$ and $\bm{\mu}_{k+1,1}^c=\bm{\mu}_{k+1}^g$.\\
For $j=1, \dots, n_{IC}$:\\
\noindent\hspace*{5mm}%
$\bm{\tilde{G}}_{k,j}=\bm{C}_{k+1,j}^c \bm{\hat{h}}_{IC,k,j}^T (\bm{\hat{h}}_{IC,k,j}  \bm{C}_{k+1,j}^c \bm{\hat{h}}_{IC,k,j}^T+ r_{IC})^{-1} ~, $    \\
\noindent\hspace*{5mm}%
 $\bm{\mu}_{k+1,j+1}^c= \bm{\mu}_{k+1,j}^c-\bm{\tilde{G}}_{k,j} \tilde{h}_{IC,j}(\bm{\mu}_{k+1}^g) ~,$ \hspace{10mm}$\inlineeqno$ \\ 
 \noindent\hspace*{5mm}%
$\bm{C}_{k+1,j+1}^c=\bm{C}_{k+1,j}^c-\bm{\tilde{G}}_{k,j} \bm{\hat{h}}_{IC,k,j} \bm{C}_{k+1,j}^c ~ .$\\
\\
As discussed in the previous subsection, the rows $\bm{\hat{h}}_{IC,k,j}$ of $\bm{\hat{H}}_{IC,k}$ are zero if the respective constraint $j$ is not active. Since an EKF update with a zero measurement vector has no effect, another exact reformulation can be applied, which leads to a dramatic speedup. \\
\\
Set $\bm{C}_{k+1,1}^c=\bm{C}_{k+1}^p$ and $\bm{\mu}_{k+1,1}^c=\bm{\mu}_{k+1}^g$. \\
Evaluate $\bm{t}_k=\bm{H}_{IC}\bm{\mu}_{k+1}^g$ with $\bm{H}_{IC}=[\bm{h}_{IC,1},\bm{h}_{IC,2},..]^T$.\\
For $j=1, \dots, n_{IC}$:\\
\noindent\hspace*{5mm}%
If $s_j (\bm{t}_k(j)-B_j)>0$:\\
\noindent\hspace*{10mm}%
$\bm{\tilde{G}}_{k,j}=\bm{C}_{k+1,j}^c \bm{h}_{IC,j} (\bm{h}_{IC,j}^T  \bm{C}_{k+1,j}^c \bm{h}_{IC,j}+ r_{IC})^{-1} ~, $    \\
\noindent\hspace*{10mm}%
 $\bm{\mu}_{k+1,j+1}^c= \bm{\mu}_{k+1,j}^c-\bm{\tilde{G}}_{k,j}  s_j( \bm{t}_k(j)-B_j) ~, $ \hspace{2mm}$\inlineeqno$ \\ 
 \noindent\hspace*{10mm}%
$\bm{C}_{k+1,j+1}^c=\bm{C}_{k+1,j}^c-\bm{\tilde{G}}_{k,j} \bm{h}_{IC,j}^T \bm{C}_{k+1,j}^c ~ .$\\
\\
Here, we also used the property $s_j^2=1$.

\subsection{Upper-Bounding the Computational Load}
\begin{figure}
	 \begin{center}
		 \includegraphics[width=1\linewidth]{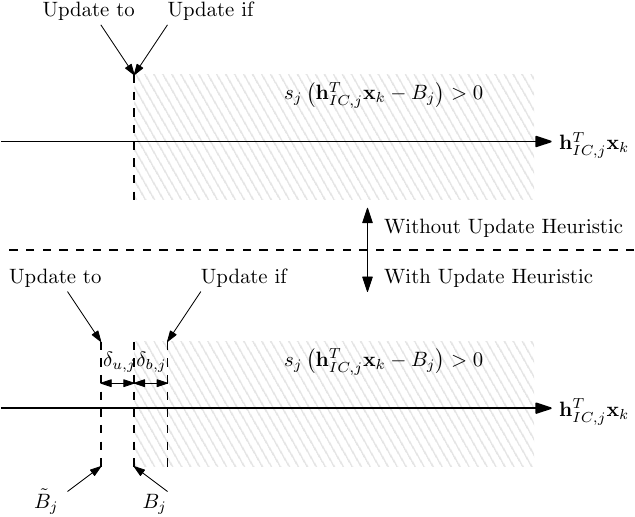}
		  \caption{Schematic illustration of the update heuristic.} 
		  \label{pic:RGP_monoton_Skizze}
	 \end{center}
\end{figure}

Since the algorithm is utilized for model learning, we can expect that our measurements, in majority, do not violate our constraints, but the constraint integration rather improves the speed at which learning takes place. To further reduce the maximum computational load per time step, we upper-bound the sequential pseudo-measurement updates per time step to $\tilde{n}_{IC}\leq n_{IC}$. Practice has shown that in cases  where the actual hidden function is close to the bounds, noise and numerical errors may lead to repetitive updates w.r.t. certain constraints. In combination with the upper-bounding of the pseudo-measurement updates per time step, this may lead to cases where certain constraints are not considered altogether. Thus, we introduce an additional heuristic to enable a hysteresis-like behavior: As illustrated in Fig. \ref{pic:RGP_monoton_Skizze}, we only update if $s_j (\bm{t}_k(j)-B_j)>\delta_{b,j}$. Moreover, we update to the value $\tilde{B}_j=B_j-\delta_{u,j} s_j$, where $\delta_{b,j}\geq0$ and $\delta_{u,j}\geq0$ are both small non-negative constants. 

\subsection{GP Monotonicity Assumptions as Constraints}
\label{sec:Monotonicity}
Monotonicity assumptions on a real, continuously differentiable function $z_k(\bm{\zeta})$ can be stated as inequality constraints of its partial derivatives w.r.t. its inputs, e.g. $\frac{\partial z_k}{\partial \zeta_i}\gtrless 0$. According to the derivation in \cite{mchutchon:2015}, the mean values of the partial derivative of a GP with SE kernels in the test grid $\bm{X}_k$ w.r.t. the input ${\zeta}_{i,k}$ is given by 
\begin{equation}
\frac{\partial \bm{\mu}_{k}^p}{\partial {\zeta}_{i,k}}= \underbrace{\left(-\frac{\beta_i}{L} \left(\left(\bm{X}_{i,k}-\bm{X}_i^T\right)\odot k(\bm{X}_k,\bm{X})\right)\cdot \bm{K}_I \right)}_{\bm{H}_{IC,l,k}} \cdot\bm{\mu}^g_{k}, \label{eq:gradient}
\end{equation}
where $\odot$ denotes the Schur- or Hadamard product. This function is linear in the mean values of the RGP. For a constant test grid $\bm{\tilde{X}}$, the linear gain matrix $\bm{H}_{IC,l,k}$ is also time-invariant. For simplicity, we only consider one monotonicity  assumption per dimension, where $l=1, \dots, n_{l}$ with $n_{l}\leq n_{\zeta}$ indicate the input dimensions for which  monotonicity assumptions exist. Consequently, each dimension is characterized by only one set of values $s_l$, $B_l$, $\delta_{B,l}$ and $\delta_{U,l}$. As a constant test grid is considered for the monotonicity, we can state the mean values of the partial derivatives at the test vector grid using linear pseudo-measurement matrices. They can be calculated by means of
\begin{equation}
\frac{\partial \bm{\mu}_{k}^p}{\partial {\zeta}_{l}}=\underbrace{\left(-\frac{\beta_l}{L} \left(\left(\bm{\tilde{X}}_{l}-\bm{X}_l^T\right)\odot k(\bm{\tilde{X}},\bm{X})\right)\cdot\bm{K}_I \right)}_{\bm{H}_{IC,l}} \cdot\bm{\mu}^g_{k} \label{eq:gradient}
\end{equation}
for all dimensions $l$ that comply with monotonicity assumptions. Here, the matrices $\bm{H}_{IC,l}$ can be calculated offline.

As confirmed in a comparison with the exact covariance of the partial derivative, see \cite{mchutchon:2015}, the usage of these measurement functions for the covariance prediction of the partial derivative leads to further linearization errors. This error vanishes if  $\bm{\tilde{X}}=\bm{X}$ holds. In the following, we therefore choose the monotonicity test vector grid equal to the basis vector grid -- which is reasonable for most applications anyway.

\subsection{Inequality Constraints on the GP Output}
\label{sec:outpuconstr}
Even though it is not employed in this paper, the consideration of inequality constraints on GP outputs is included here for completeness. Like the monotonicity constraints, output constraints are evaluated on a grid. If the chosen test grid is also identical to the basis vector grid, the corresponding output equation becomes $\bm{\mu}_{k+1}^p= \bm{J}_{k}(\bm{X}) \bm{\mu}_{k+1}^g$. Since $\bm{J}_{k}(\bm{X})=k(\bm{X},\bm{X})  \bm{K}_I=\bm{I}$ holds, the respective pseudo-measurement matrix becomes the identity matrix $\bm{H}_{IC}=\bm{I}$ and leads to the simplification $\bm{t}_k=\bm{\mu}_{k+1}^g$. Please note that for soft GP output constraints, other measures, like a bounding of the measurements $y_k$ or the RGP mean values $\bm{\mu}^g_{k}$, might be more efficient and effective.

\subsection{Summary of GP Monotonicity Constraints}
\label{sec:summary}
This section summarizes the pseudo-measurement updates for constraints related to monotonicity assumptions in the RGP. The extended RGP algorithm will in the following referred to as RGPm. Please note that the index $k$ denoting the time step is omitted for simplicity. 
In addition to the offline precomputations (2) for the standard RGP in Sec.~\ref{sec:RGP}, the following measurement matrices need to be calculated for each input dimension with a monotonicity assumption: \\
\\
\noindent%
$\mathrm{For} ~l =1, \dots, n_{l}$: \\
\noindent\hspace*{5mm}%
$ \bm{H}_{IC,l}= \left(-\frac{\beta_l}{L} \left(\left(\bm{X}_l-\bm{X}_l^T\right)\odot \bm{K}\right)\cdot\bm{K}_I  \right)\,.$ \hspace{5mm}$\inlineeqno$ \\
\\
Also the boundaries $\tilde{B}_l=B_l-\delta_{u,l} s_l$ and $\delta_{b,l}$ for the update heuristic as well as the upper limits per dimension $\tilde{n}_{IC,l}$ have to be defined.

For each time step $k$, after  the evaluation of the RGP  inference (3) and update steps (4) from Sec.~\ref{sec:RGP},  the pseudo-measurement update is computed as follows:     \\
\\
Set $\bm{C}_{1,1}^c=\bm{C}_{k+1}^g$ and $\bm{\mu}_{1,1}^c=\bm{\mu}_{k+1}^g$.\\
$\mathrm{For}~ l=1, \dots, n_{l}$: \\
\noindent\hspace*{5mm}%
Evaluate $\bm{t}_{l}=\bm{H}_{IC,l}\bm{C}_{l,1}^c$, with $\bm{H}_{IC,l}=[\bm{h}_{IC,l,1},\bm{h}_{IC,l,2},..]^T$ and set counter $m=1$. \\
\noindent\hspace*{5mm}%
$\mathrm{For} ~j=1, \dots, n_{IC,l}$: \\
\noindent\hspace*{10mm}%
If $s_l (\bm{t}_{l}(j)-\tilde{B}_l)>\delta_{b,l}$ and $m\leq \tilde{n}_{IC,l}$: $\hspace{0mm}$  $~,\inlineeqno$ \\
\noindent\hspace*{15mm}%
$\bm{\tilde{G}}_{l,j}=\bm{C}_{l,j}^c \bm{h}_{IC,l,j} (\bm{h}_{IC,l,j}^T  \bm{C}_{l,j}^c \bm{h}_{IC,l,j}+ r_{IC})^{-1} $    \\
\noindent\hspace*{15mm}%
 $\bm{\mu}_{l,j+1}^c= \bm{\mu}_{l,j}^c-\bm{\tilde{G}}_{l,j} s_l (\bm{h}_{IC,l,j} \bm{\mu}_{l,1}^c-\tilde{B}_l) $ \\
 \noindent\hspace*{15mm}%
$\bm{C}_{l,j+1}^c=\bm{C}_{l,j}^c- \bm{\tilde{G}}_{l,j} \bm{h}_{IC,l,j} \bm{C}_{l,j}^c $ \\
 \noindent\hspace*{15mm}%
$m=m+1$\\
 \noindent\hspace*{5mm}%
$\bm{\mu}_{l+1,1}^g=\bm{\mu}_{l,\tilde{n}_{IC,l}}^c$ and $\bm{C}_{l+1,1}^c=\bm{C}_{l,\tilde{n}_{IC,l}}^c$ \\
Set $\bm{\mu}_{k+1}^g=\bm{\mu}_{n_{l},\tilde{n}_{IC,l}}^c$, but reset $\bm{C}_{k+1}^g=\bm{C}_{1,1}^c$.

\subsection{Discussion}
\label{sec:discussion}
In general, a consideration of monotonicity constraints with a test vector grid unequal to the basis vector grid is possible and was successfully implemented. It tends to be, however, numerically unstable, most likely due to linearization errors w.r.t.~the covariance prediction.

As pointed out, the implemented monotonicity constraints are only soft constraints. This means in practice that the GP will adhere to the measurements if these consistently contradict the monotonicity assumptions. This scenario would be caused by wrong assumptions regarding either the monotonicity itself or the measurement quality.  Although still not guaranteeing constraint satisfaction, a safety margin can be added by changing $B_l$ accordingly. For safety-critical constraints, an additional evaluation on a finer grid as well as a subsequent output correction may be useful.

A UKF may be more suited for handling the nonlinearities in the presented pseudo-measurement update. Nevertheless, it would prevent most of the reformulations which facilitate the envisaged real-time implementation. Consequently, it was not considered further.

Please note that piece-wise monotonicity can also be addressed as long as the regions are definable through sets of basis vector points. This variation and the direct output constraints in Subsec.~\ref{sec:outpuconstr} have already been successfully validated but are not described for the sake of brevity. 

Please note that the sequential update of the EKF involves an additional condition for the equivalence to a batchwise update. Here, the linearization point has to be identical. This condition is fulfilled within each input dimension by introducing the intermediary variables $\bm{t}_k$ and  $\bm{t}_l$ in Subsecs.~\ref{sec:seq_EKF} and \ref{sec:Monotonicity}, respectively. It is, however, not fulfilled over all input dimensions $l$ in Subsec.~\ref{sec:Monotonicity}. Thus, even without the bounded number of updates and the update heuristic, a slight deviation is present between the sequential EKF and the batchwise EKF if several monotonicity assumptions apply.

\section{Validation Based on Simulations}
\label{sec:simval}
As a first validation example, the hidden static function $z_k=2 (1+0.1 \zeta_k+\zeta_k^3) $ is learned. The input $\zeta_k$ is picked from a random uniform distribution over the complete input range $\zeta_k \in [-1,1]$. Zero-mean Gaussian white noise with a variance of  $\sigma_y^2=1e-2$ is added to the measured output $y_k$.

The RGP hyperparameters remain constant and are chosen as follows: $L=3$, $\sigma_K=1e1$, and $N=21$. Obviously, the hidden function is strictly monotonically increasing, so $\frac{\partial z}{\partial \zeta}>0$,  $B_1=0$, and $s_1=-1$ hold. The pseudo-measurement noise is chosen as $r_{IC}=1e-8$.

\subsection{Simulation Results}
\begin{figure}
	 \begin{center}
		 \includegraphics[width=1\linewidth]{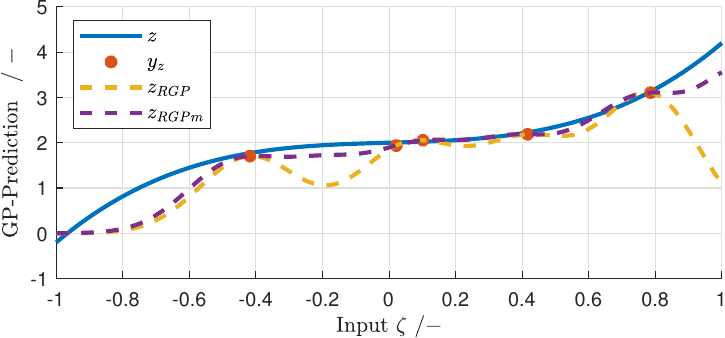}
		  \caption{RGP and RGPm outputs in comparison with the hidden function $z$ after 5 time steps.} 
		  \label{pic:BSP_mean}
	 \end{center}
\end{figure}

In Fig. \ref{pic:BSP_mean}, the hidden function $z$ and its learned reconstructions $z_{RGPm}$ and $z_{RGP}$ -- with and without monotonicity assumptions -- are depicted after five noisy measurements $y_k$. It becomes obvious that the RGPm algorithm works properly. In this example, it is capable of enforcing strict monotonicity over the whole input range, while the curve is not significantly altered in the vicinity of the actual measurement points. The result represents a clearly better fit in comparison with the classical RGP -- especially in regions where still no measurements are available.

For a statistical validation of the algorithm and an assessment of the impact of the adaptations, we investigate six different variants of the algorithm:
\begin{itemize}
\item[S0] Pure RGP
\item[S1] unlimited  $\tilde{n}_{IC}=n_{IC}$ RGPm with  $\delta_b=0$ and $\delta_u=0$ 
\item[S2] $\tilde{n}_{IC}=2$ limited RGPm with  $\delta_b=0$ and $\delta_u=0$  
\item[S3] $\tilde{n}_{IC}=2$ limited RGPm with  boundary heuristic $\delta_b=1e-1$ and $\delta_u=1e-1$ 
\item[S4] $\tilde{n}_{IC}=5$ limited RGPm with  $\delta_b=0$ and $\delta_u=0$  
\item[S5] $\tilde{n}_{IC}=5$ limited RGPm with  boundary heuristic $\delta_b=1e-1$ and $\delta_u=1e-1$ 
\end{itemize}

\begin{figure}
	 \begin{center}
		 \includegraphics[width=1\linewidth]{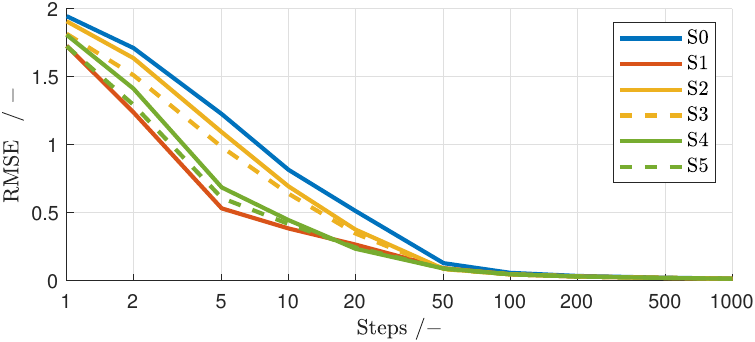}
		  \caption{Average RMSE for 500 simulation runs of the different RGP and RGPm variants.} 
		  \label{pic:Sim_Stat_RSME}
	 \end{center}
\end{figure}

\begin{figure}
	 \begin{center}
		 \includegraphics[width=1\linewidth]{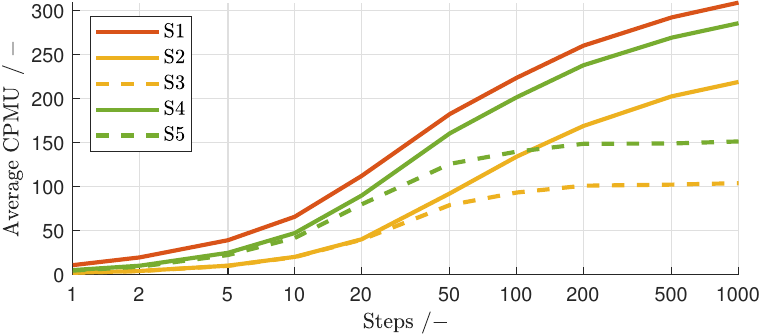}
		  \caption{Average cumulative pseudo-measurement updates (CPMU) for 500 simulation runs of the different RGPm variants.} 
		  \label{pic:Sim_Stat_cU}
	 \end{center}
\end{figure}

For each scenario, 500 simulation runs are conducted with the described uniform random input and the noisy output. After $k=1,2,5, \dots ,1000$ steps, the root mean-squares error (RMSE) between the learned function of each variation, compared to the hidden function over the complete input range, is calculated. This RMSE is again averaged over all 500 simulations and depicted in Fig. \ref{pic:Sim_Stat_RSME}. Fig.~\ref{pic:Sim_Stat_cU} shows the average cumulative number of pseudo-measurement updates (CPMU) that were conducted for the  alternatives S1,...,S5. 

All RGPm variants perform better than the classical RGP. This advantage vanishes after a certain number of steps, and all the RMSE converge to the same value. This could be expected because the monotonicity assumptions are valid and should be represented on average in the measurements. 
 
It can be seen that a tighter limit on the number of pseudo-measurement updates per time step is generally a trade-off w.r.t.~the performance. Accordingly, S1 is better than S4 and S2. In the current application, the improvement from S4 to S1 is however quite small, and reverses slightly around 20 steps. This last effect may stem from numerical errors that could be related to the higher number of pseudo-measurement updates.

The heuristic approach of including outer/inner boundaries resulted in significant improvements in the RMSE. The reduced repeated pseudo-measurement updates w.r.t. certain constraints are clearly visible in Fig.~\ref{pic:Sim_Stat_cU}. While S1, S2 and S4 still perform pseudo-measurement updates after 1000 steps, the update heuristic leads to a converging CPMU for S3 and S5, which means that no more pseudo-measurement updates are performed after a certain point. Whereas the real-time capability is only influenced by the maximum number of updates per time step, the reduced computational load can be seen as an additional benefit.

\section{Experimental Results for a Vapor Compression Cycle}
\label{sec:expval}
In this chapter, the functionality of the proposed method is investigated in a real-time application -- a counterflow evaporator as a subsystem of a VCC. As  discussed in detail in \cite{VCC_MB:2024} and \cite{VCC_Part:2024}, heat transfer values, especially in the presence of phase changes, depend on process parameters and are hard to determine. In \cite{VCC_Part:2024}, integral feedback was applied directly affecting the heat transfer value and combined with a partial IOL. In addition to a stability proof also the convergence of  integrator state towards the actual heat transfer value was proven. In this paper we want to leverage the integrator steady state to learn a RGP model for the heat transfer value that depends on several process parameters. The aim is to use the RGP model in parallel to the integrator feedback to further improve the transient control behavior.

The algorithm was implemented on a test rig at the Chair of Mechatronics at the University of Rostock. The utilized hardware, a Bachmann PLC (CPU:MH230), runs at a sampling time of $10ms$. The test rig is fully equipped with sensors, an electronic expansion valve as well as a compressor with a variable frequency drive. Please refer to \cite{VCC_MB:2024} and \cite{VCC_Part:2024} for further details about the test rig.

\subsection{Partial IOL Control}
\begin{figure}
	 \begin{center}
		 \includegraphics[width=1\linewidth]{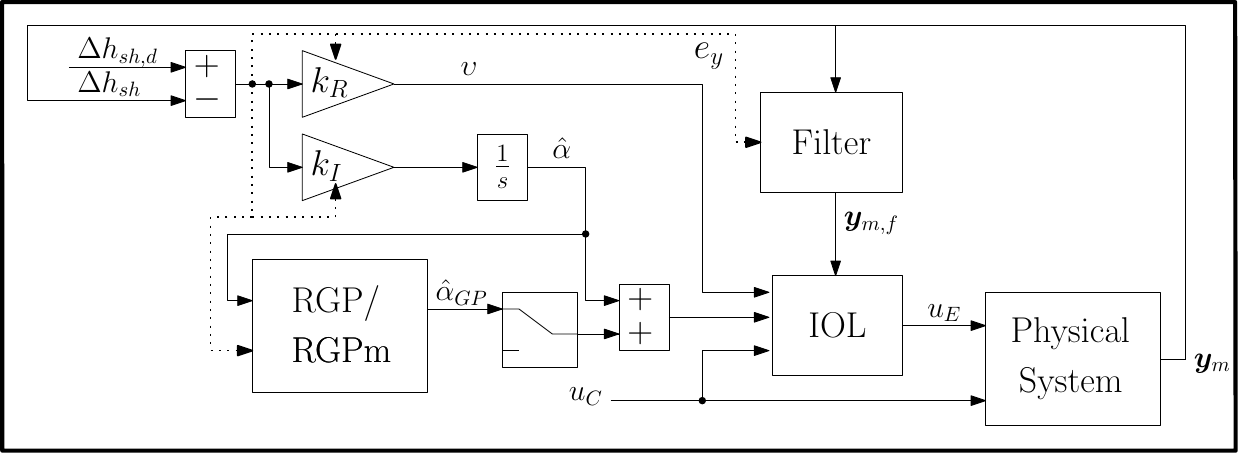}
		  \caption{Nonlinear tracking control structure.} 
		  \label{pic:ControlStructure}
	 \end{center}
\end{figure}

As the partial IOL is described in detail in \cite{VCC_Part:2024}, we only want to give a brief overview in this paper. The envisioned control structure is depicted in Fig.~\ref{pic:ControlStructure}. As shown, the control inputs are given by the expansion valve opening $u_E$ and the compressor speed $u_C$. The control outputs  are chosen as the evaporator mean temperature $y_1=T_{Evap}$ and the superheating enthalpy $y_2=\Delta h_{SH}$ after the evaporator.  The control-oriented evaporator model, first derived in \cite{VCC_IOL:2023}, is governed by the following two ODEs 
\begin{equation}
\resizebox{1\columnwidth}{!}{$
\dot{h}_{Evap}=\frac{\dot{m}_{E}  h_{Evap,in}-\dot{m}_{C}  h_{Evap,out}+\dot{Q}_{Evap}-(\dot{m}_{E}-\dot{m}_{C})h_{Evap} }{V_{Evap} \rho_{Evap}}  \label{eq:hEv}$
}
\end{equation}
and
\begin{equation}
\dot{\rho}_{Evap}=\frac{1}{V_{Evap}} \left(\dot{m}_{E}-\dot{m}_{C} \right)~. \label{eq:dEv}
\end{equation}
Here, the first state equation is affected by an uncertain convective heat flow
\begin{equation}
\dot{Q}_{Evap}=\alpha \cdot \alpha_{norm}\cdot A_{Evap} \cdot (T_U-T_{Evap})\,
\end{equation} 
with $\alpha_{norm}$ as a normalization factor. The input-dependent mass flows are given by
\begin{equation}
\dot{m}_E=\zeta_E \cdot \sqrt{\rho_{Cond,out} (p_{Cond}-p_{Evap})} \cdot {u_{E}}
\end{equation}
and
\begin{equation}
\dot{m}_C=\hat{m}_{Rev} u_C \eta_V \,.
\end{equation}

Both outputs can be expressed in terms of the state variables. The second output is given by
\begin{equation}
\resizebox{1\columnwidth}{!}{$
{y}_{lin,2}=\Delta h_{SH} = \frac{h_{Evap}-(1-\mathrm{w_\rho})h_{Evap,in}}{\mathrm{w_\rho}}  -h_{Evap,sat}(T_{Evap})\,,$
}
\end{equation}
and its first time derivative becomes
\begin{equation}
\resizebox{1\columnwidth}{!}{$
\dot{y}_{lin,2}=\frac{1}{\mathrm{w_\rho}}  \dot{h}_{Evap}-\frac{1-\mathrm{w_\rho}}{\mathrm{w_\rho}} \frac{d h_{Evap,in}}{dt}-\left.\frac{\partial h_{sat}}{\partial T}\right|_{{T}_{Evap}} \dot{T}_{Evap}\,,$
}
 \label{eq:dyfl2}
\end{equation}
which lead to a relative degree of one. Whereas the inverse dynamics is determined for both control inputs in the standard IOL proposed in \cite{VCC_IOL:2023}, the partial IOL used here leverages the inverse dynamics of the control input $u_E$ 
\begin{equation}
	 u_E = b_3(.)^{-1} \left[\upsilon_2-a_2(.)-b_4(.) u_C\right] \,,
\end{equation}
and considers the second input $u_C$ as a measurable disturbance. 
Please note that a secondary controller needs to be implemented for the first control output $y_1=T_{Evap}$, see \cite{VCC_Part:2024}. In this paper, however, the control input $u_C$ is assumed as given because it has no relevance for the presented algorithm. 
The partial IOL is completed with a stabilizing feedback $\upsilon_2=k_R(e_y)e_y$, with $e_y=y_{2,d}-y_2$. Here the nonlinear feedback from \cite{VCC_Part:2024} is utilized. The introduction of a parameter update law for the heat transfer value $\dot{\hat{\alpha}}=-k_I (y_{2,d}-y_2)$, which corresponds to integral control action w.r.t.~$\hat{\alpha}$, dramatically increases the performance and guarantees steady-state accuracy as shown in \cite{VCC_Part:2024}.
Since the integrator state $\hat{\alpha}$ was proven to converge to an exact representation of the lumped heat transfer value in the control-oriented model, it is suitable for the training of a model.

\subsection{RGPm Implementation}
As input variables for the GP model, the compressor speed $u_C$ and the superheating enthalpy $y_2=\Delta h_{SH}$ are chosen, i.e., $\hat{\alpha}_{GP}=z(u_C,\Delta h_{SH})$. As known from our own experiments and from the literature, the inequalities $\frac{\partial \alpha}{\partial u_C}>0$  and $\frac{\partial \alpha}{\partial \Delta h_{SH}}<0$ hold. Consequently, they are included as monotonicity assumptions in the RGPm algorithm. The evaluation of the GP model is conducted with the reference value $y_{d,2}$ instead of the current one, i.e., $\hat{\alpha}_{GP}=z(u_C,\Delta h_{SH,d})$. Thereby, any additional feedback is avoided, and the GP model obtains a pure feedforward structure w.r.t. its evaluation. This prevents stability issues.

The RGP and RGPm algorithms are implemented with basis vector dimensions of $N_1=5$ and $N_2=5$ for the respective GP inputs, which results in $N_1 N_2=25$ basis vector points. The monotonicity update is bounded to $n_1=n_2=5$ steps per dimension and per time step. The parameters for the update heuristics are defined as follows: $\delta_{u,1}=\delta_{u,2}=0$ and $\delta_{b,1}=\delta_{b,2}= 2e-6 $. The pseudo-measurement noise is chosen as $r_1=r_2=1e-8$, and the length scale is set to $L=1.5$. The signing variables for the respective monotonicity assumptions are given by $s_1=-1$ and $s_2=1$ and the safety margins are disabled, so $B_1=B_2=0$. 

To ensure that the integrator state is only utilized in the vicinity of it's equilibrium state, we employ a heuristic which only enables the RGP update if $|e_y|<1$ holds for $10$ consecutive seconds. The numerical values in this heuristic mark a trade-off between learning speed and accuracy. With a larger error margin or a shorter time horizon, more data qualifies for training. However, the considered data might also relate to system states further away from the equilibrium.

\subsection{Results}
\begin{figure*}[h]
	 \begin{center}
\includegraphics[width=0.325\linewidth]{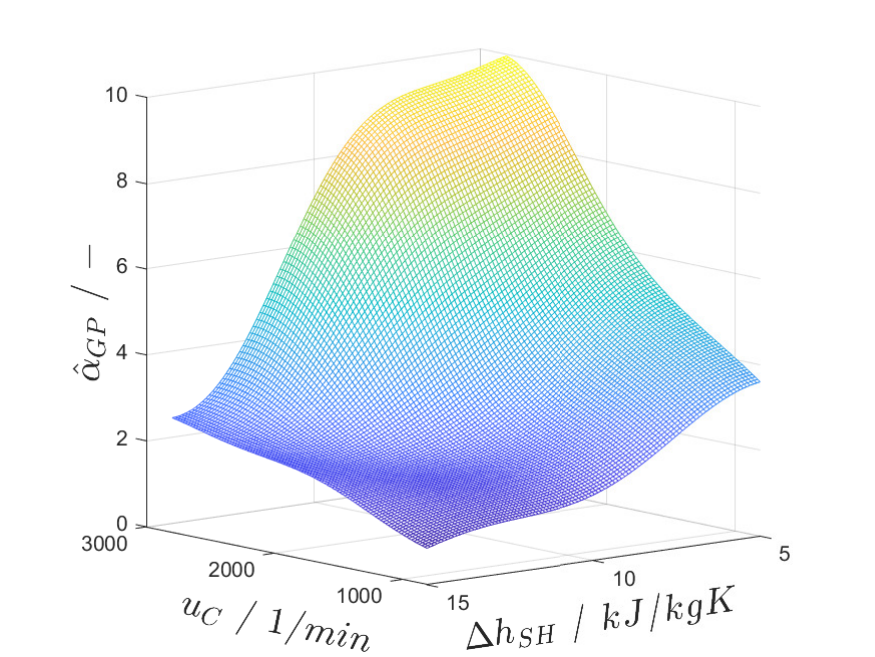} 
\includegraphics[width=0.325\linewidth]{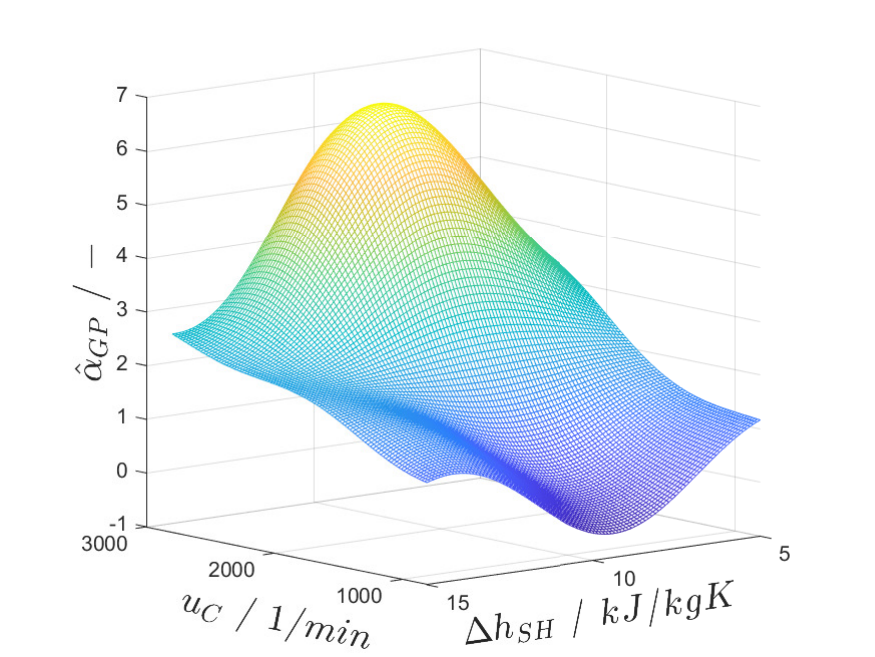} 
\includegraphics[width=0.325\linewidth]{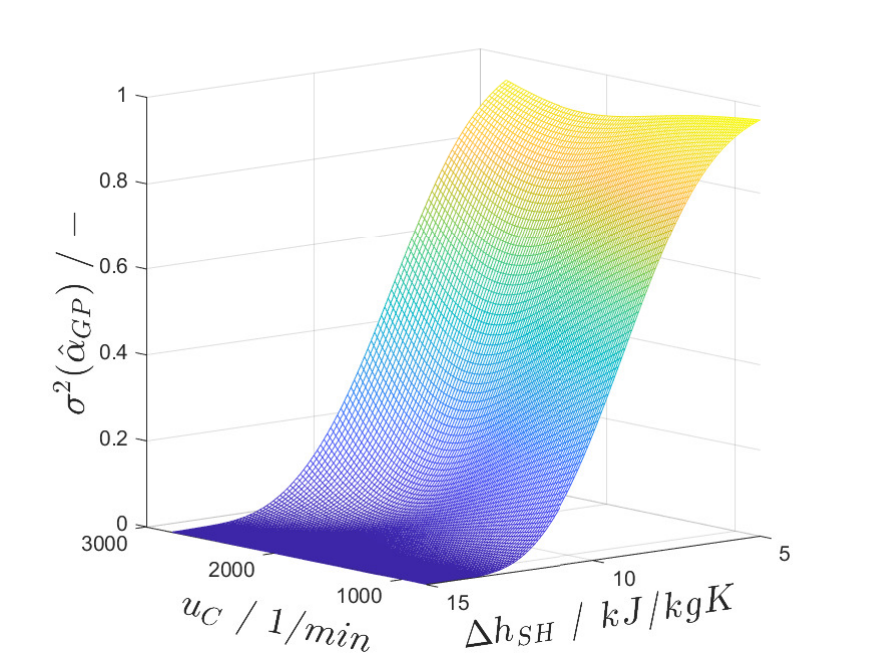} 
\includegraphics[width=0.325\linewidth]{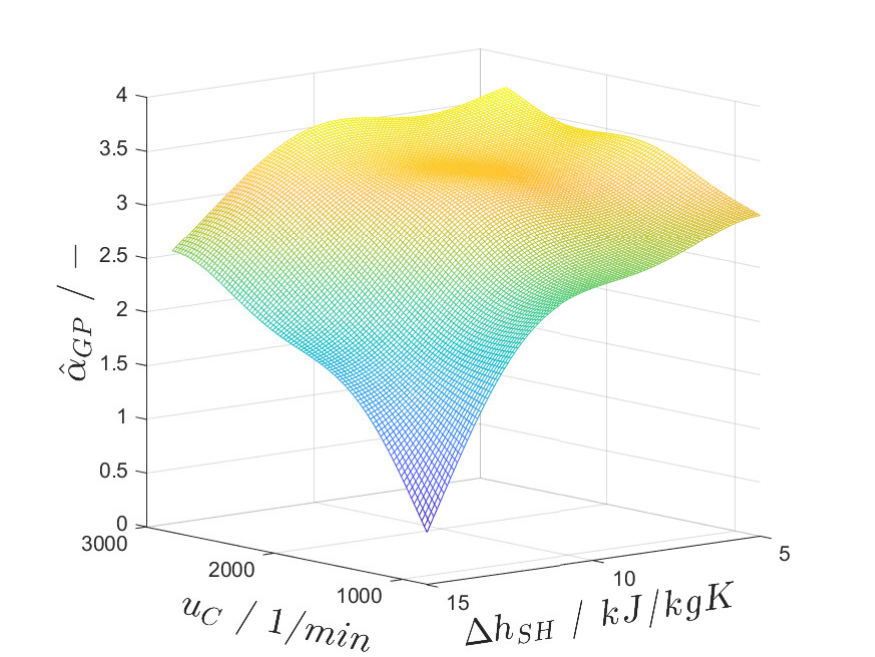}
\includegraphics[width=0.325\linewidth]{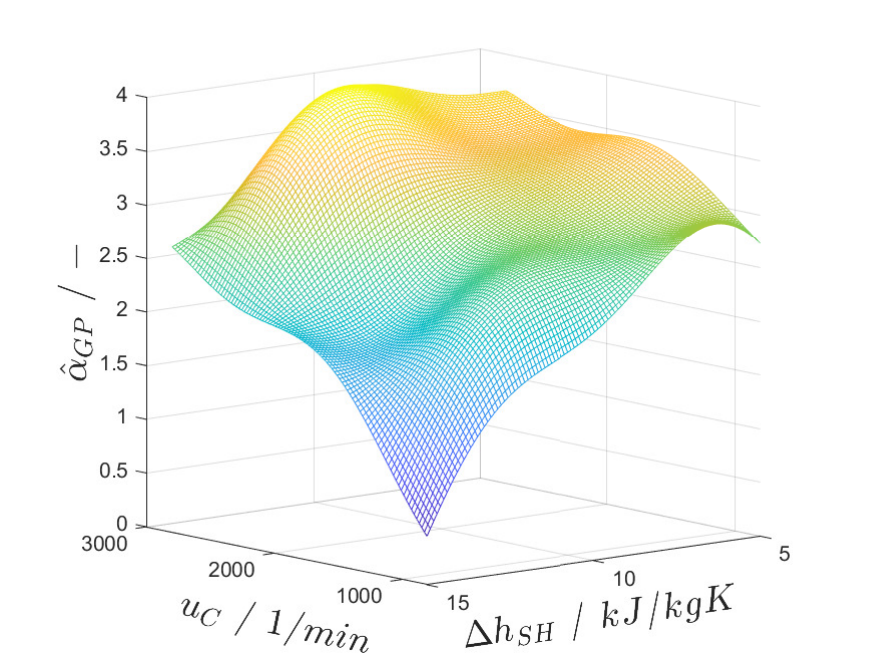}
\includegraphics[width=0.325\linewidth]{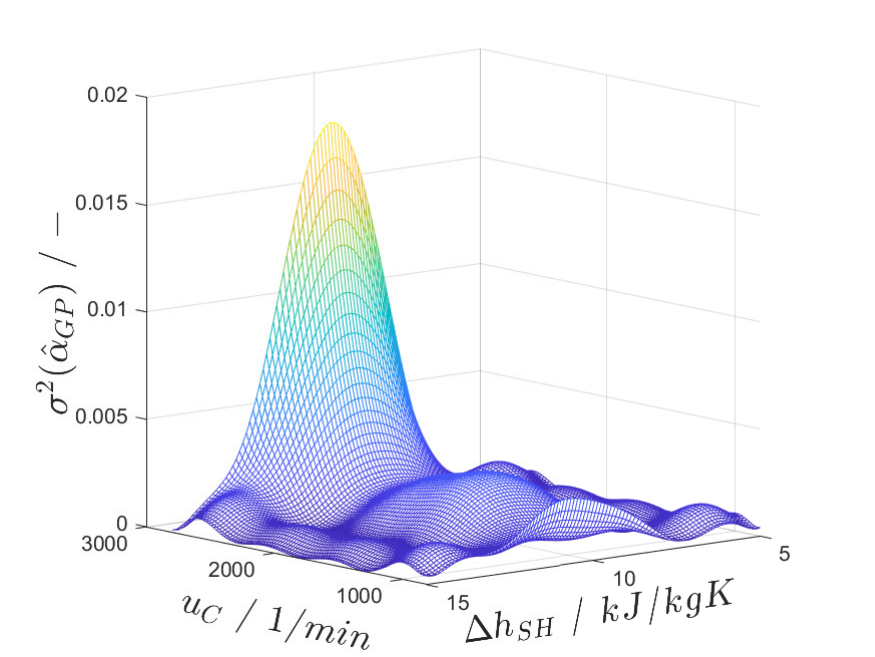}
		  \caption{GP model predictions from left to right: RGPm (online), RGP (offline comparison for same data) and covariances of both variants.  From top to bottom: after the first $u_C$-trajectory for $y_{2,d}=14$, and after the second $u_C$-trajectory for $y_{2,d}=7$.} 
		  \label{pic:alpha_fac_sol}
	 \end{center}
\end{figure*}

To illustrate the functionality and advantages of the RGPm algorithm, two consecutive experiments with the same step-wise $u_C$-trajectory are performed: the first one for $y_{2,d}=14$, the other one for $y_{2,d}=7$. The RGPm algorithm is trained online during both experiments -- without applying its output value to the controller. For a comparison, the RGP algorithm is later trained offline on the same data. The results after the first and second experiment are depicted for both algorithms in Fig.~\ref{pic:alpha_fac_sol}. Since the monotonicity updates for the covariance matrix are discarded at the end of each time step, both algorithms lead to the same covariance, so that only one is depicted per experiment. In the upper right picture of Fig.~\ref{pic:alpha_fac_sol}, the training region during the experiment becomes obvious, because the covariance is small there. When comparing the mean values for the RGPm (upper left) and RGP (upper middle), it becomes evident that the learned model within the input region used for the training is nearly identical. Outside of this region, however, the shapes vary drastically. While the monotonicity assumptions in the RGPm model lead to a physically viable model shape, this cannot be stated for the standard RGP model. Here, the overall shape is only defined by the limited amount of data available in the non-trained region. As shown in the comparison of the trained models after the second experiment for both RGPm (lower left) and RGP (lower middle), the differences between the models decline with the availability of data. This reinforces the findings from the simulation study.

The softness of the constraints in the RGPm algorithm becomes clear for low superheating and large compressor speeds. Here, the monotonicity constraints are slightly violated, however much less than in the standard RGP algorithm. The violation may be fixed by safety margins and a more involved heuristic for the RGP update.

The evaluation of the overall control performance with the trained model is outside of the scope of this paper, since a more detailed comparison would need additional concepts and algorithms, e.g. suitable training strategies. A quick validation of the RGPm model after the second experiment, depicted in Fig. \ref{pic:alpha_fac_sol} (lower left corner), can be performed by deactivating the integrator feedback control and running a $u_C$ test trajectory together with a learned $\hat{\alpha}_{GP}$ and with a constant $\alpha$. Here, for a reference value of $y_d=13$, which wasn't used for the training, a $28~\%$ reduction in RMSE could be achieved, which represents a promising first result for the application of the trained RGPm model.

\section{Conclusions and Outlook}
This paper presents an extension of the recursive Gaussian Process regression (RGP) algorithm to enforce (soft) constraints during an online training. Here, the consideration of monotonicity assumptions in the GP model is emphasized as well as their implementation as constraints. The algorithm is statistically validated in simulations. Moreover, different simplifications are introduced to enable a real-time use, and their effect on the performance is investigated. A real-time implementation on a test rig is presented afterwards, where the learning of heat transfer value models for an evaporator is considered. The evaporator is a main component of a vapor compression cycle operated by a previously published partial IOL. Here, the extension of a standard RGP regression by monotonicity constraints results in a significantly better modeling -- especially if few data is available. 

A combination of the extension presented in this paper with the Kalman Filter integration of the RGP (KF-dRGP), is straightforward and allows for an RGP training with constraints if the output of the hidden function is not directly measurable. For systems where a hard constraint satisfaction is crucial, further investigations are still necessary. As discussed earlier, an additional safety step with a calculation of the constraint satisfaction on a finer grid could be suitable.
As mentioned in Sec.~\ref{sec:expval}, the full integration of the GP model within the partial IOL for the VCC control necessitates further work. It is obviously desirable, for example, to simultaneously learn and apply the learned model. This does, however, introduce an additional feedback loop into the system, which may affect the stability. Here, the presented output and monotonicity bounds mark important tools. Given promising first results, steps in this direction will be topics of future work.

%

\newpage

\bibliographystyle{apalike}
{\small
\bibliography{../myrefs}}

%

\end{document}